\documentclass[12pt, twoside]{article}
\usepackage{a4wide}
\usepackage{latexsym}
\usepackage[centertags]{amsmath}
\numberwithin{equation}{section}
\begin{document}
\begin{titlepage}
\begin{center}
\hspace*{10cm} BONN-TH-99-16\\
\hspace{10cm} hep-th/9909144\\
\hspace{10cm} September 1999\\
\vspace*{2cm}

{\bf \LARGE Gauged N=2 Off-Shell Supergravity in Five Dimensions}\\
\vspace{1cm}

{\sc Max Zucker}\footnote{e-mail: zucker@th.physik.uni-bonn.de}
\vspace{.2cm}

{\it { Physikalisches Institut \\
Universit\"at Bonn\\
Nussallee 12\\
D-53115 Bonn, Germany}}\\
 \end{center}
\vspace{1.2cm}

\vspace{.3cm}
\begin{center}
{\large Abstract}
\end{center}
\vspace{.3cm}

We present some multiplets of $N=2$ off-shell supergravity in five dimensions. One is the Super Yang-Mills multiplet, another one is the linear multiplet. The latter one is used to establish a general action formula from which we derive an action for the Super Yang-Mills multiplet. The Super Yang-Mills multiplet is used to construct the nonlinear multiplet with gauged $SU(2)$. This nonlinear multiplet and the action formula for the Yang-Mills multiplet enable us to write down an $SU(2)$ gauged supergravity which we finally truncate to arrive at gauged supergravity with gauge group $SO(2)$.  
\end{titlepage}
\section{Introduction}
In a recent paper \cite{Zucker:1999ej} we have presented an off-shell formulation of $N=2$ supergravity in five spacetime dimensions. Our construction started from globally supersymmetric Yang-Mills theory from which we derived the multiplet of currents. By dualizing this multiplet we obtained the linearized transformation rules of the minimal multiplet which we subsequently completed to the full, nonlinear ones. The field content of this multiplet is listed in table \ref{min}.
\begin{table}[h]
 \begin{center}
 \begin{tabular}{ccccccc}
 \hline\hline
 Type & Field & $SU(2)$ & dimension & \multicolumn{3}{c}{components}\\
&&&&bosonic&&fermionic\\

 \hline
 vielbein & $e_m^a$ & \bf 1 & 0 & $10$&& \\
 graviphoton &  $A_m$ & \bf 1 & 3/2 & $4$&& \\
 gravitino & $\psi_m$ & \bf 2 & 2 &&& $32$ \\
 isotriplet & $\vec{t}$ & \bf 3 & 5/2 & 3&& \\
 antisymmetric tensor & $v_{ab}$ & \bf 1 & 5/2 & 10&&\\
 $SU(2)$ gauge field & $\vec{V}_m$ & \bf 3 & 5/2 & $12$&&\\
 spinor & $\lambda$ & \bf 2 & 3 &&& 8\\
 scalar & $C$ & \bf 1 & 7/2 & 1&&\\
 \hline\hline
 &&&&40&+&40\\
 \end{tabular}
\end{center}
\caption{Field content of the minimal multiplet in $D=5$.}\label{min}
\end{table}
The propagating fields are the vielbein $e_m^a$, the gravitino $\psi_m$ and the graviphoton $A_m$. Besides these fields some auxilliary fields are needed which ensure the matching of bosonic and fermionic degrees of freedom off the mass shell. As can be seen from the table, the off-shell extension requires all in all $40+40$ components. Since at least $48+48$ components are needed in order to construct a supergravity lagrangian, $8+8$ components are missing. These $8+8$ components were added to the theory by introducing the nonlinear multiplet. Furthermore, using this multiplet, we were able to break the $SU(2)$ automorphisms which were gauged by the auxilliary field $\vec{V}_m$ to a global $SU(2)$, leading to an off-shell lagrangian of supergravity.

In this note we continue the construction of locally supersymmetric off-shell theories in $D=5$. We use the minimal formulation throughout, i.e. with unbroken $SU(2)$ and $40+40$ components. At the end, the nonlinear multiplet is introduced leading to the desired lagrangians.
Our final result will be gauged supergravity, i.e. {\it AdS} supergravity (For a recent review on supersymmetry in {\it AdS} space, see \cite{deWit:1999ui}). In principle there are two possible versions of gauged supergravities: 

The first one consists in gauging the complete automorphism group by introducing an additional Super Yang-Mills multiplet with gauge group $SU(2)'$. The resulting $SU(2)\times SU(2)'$ is then broken to a residual $SU(2)$ which gauges the complete automorphisms, where in contradistinction to the minimal case, the gauge field is now propagating. We will not present the explicit lagrangian but explain in some detail its construction.

The second formulation of gauged supergravity uses the graviphoton $A_m$ belonging to the supergravity multiplet to gauge an $SO(2)$ subgroup of the automorphism group. The resulting theory contains a cosmological constant. The lagrangian will be presented in section \ref{sec4}.

The plan for this note is as follows. In the next section we find the transformation laws of the locally supersymmetric Yang-Mills multiplet. In section \ref{sec3} the linear multiplet is presented. It allows the construction of a general action formula which is used to find an action for the Super Yang-Mills multiplet. In section \ref{sec4} we move on to the main topic of this work, the construction of gauged supergravity. This requires the inclusion of a nonabelian gauge symmetry in the nonlinear multiplet is presented there. Finally, we give the lagrangian for off-shell supergravity in Anti de Sitter space.

\section{The Super Yang-Mills Multiplet}
The field content of the Super Yang-Mills multiplet is given in table \ref{table1}.
\begin{table}[t]
 \begin{center}
 \begin{tabular}{ccccccc}
 \hline\hline
 Type & Field & $SU(2)$ & dimension & \multicolumn{3}{c}{components}\\
&&&&bosonic&&fermionic\\
 \hline
 gauge field & $B_m^A$ & \bf 1 & 3/2 & $5-1$&& \\
 scalar &  $M^A$ & \bf 1 & 3/2 & 1&& \\
 gaugino & $\Omega^A$ & \bf 2 & 2 & && 8  \\
 isotriplet & $\vec{X}^A$ & \bf 3 & 5/2 & 3&& \\
 \hline\hline
 &&&&8&+&8\\
 \end{tabular}
 \end{center}
 \caption{Field content of the Super Yang-Mills multiplet in $D=5$. The index $A$ is a gauge group index.}\label{table1}
 \end{table}
 We found the locally supersymmetric transformation laws starting from the rigid ones which are given in \cite{Mirabelli:1998aj} by imposing the commutator algebra of the minimal multiplet, i.e. the multiplet containing 40+40 components as constructed in our recent work \cite{Zucker:1999ej}. One finds that this algebra has to be modified. In addition to the terms given there, a field dependent gauge transformation appears in the commutator of two supersymmetry transformations:
\begin{equation}
[\delta_Q(\eta),\delta_Q(\varepsilon)]=\ldots +\delta_{gauge}(ig\bar{\varepsilon}\gamma^m\eta B_m^A-ig\bar{\varepsilon}\eta M^A).\label{alg2}
\end{equation}
The dots represent the field dependent transformations of \cite{Zucker:1999ej}, $\delta_{gauge}(\alpha)$ is a gauge transformation with parameter $\alpha$ and $g$ is the gauge coupling constant which has (mass-) dimension $-1/2$ in five dimensions. 

The tranformation laws we find are
\begin{equation}
\begin{split}
\delta B_m^A &  = i\bar{\varepsilon}\gamma_m\Omega^A-i\bar{\varepsilon}\psi_mM^A \\
\delta M^A & = i\bar{\varepsilon}\Omega^A\\
\delta \Omega^A & =  \frac{1}{4}\gamma^{ab}\varepsilon\widehat{G}_{ab}^A-\frac{1}{2}\gamma^a\varepsilon\widehat{\cal D}_aM^A + i\vec{X}^A\vec{\tau}\varepsilon-\frac{1}{2\sqrt{3}}\gamma^{ab}\varepsilon\widehat{F}_{ab}M^A
\\
\delta \vec{X}^A & = -\frac{1}{2}\bar{\varepsilon}\vec{\tau}\gamma^a\widehat{\cal D}_a\Omega^A-\frac{g}{2}f_{BC}{}^{A}\bar{\varepsilon}\vec{\tau}\Omega^B M^C +\frac{1}{4}\bar{\varepsilon}\vec{\tau}\gamma^{ab}\Omega^A v_{ab}+4i\bar{\varepsilon}\Omega^A\vec{t}-i\bar{\varepsilon}\vec{\tau}\tilde{\tau}\Omega^A\tilde{t},
\end{split}\label{sym}
\end{equation}
where $f_{AB}{}^{C}$ are the structure constants of the Lie algebra
\[
[T_A,T_B]=if_{AB}{}^{C}T_C.
\]
Our notations and conventions are the same as in our previous work \cite{Zucker:1999ej}, except for the gravitational coupling $\kappa$ which is now set to one. For the convenience of the reader we list the explicit expressions for the supercovariant objects as they appear in the transformation laws (\ref{sym}):
\begin{eqnarray*}
\widehat{\cal D}_m M^A & = & \partial_m M^A+gf_{BC}{}^A B_m^B M^C+i\bar{\Omega}^A\psi_m\\
\widehat{G}_{mn}^A & = & \partial_m B_n^A-\partial_n B_m^A + gf_{BC}{}^{A} B_m^BB_n^C-2i\bar{\psi}_{[m}\gamma_{n]}\Omega^A+i\bar{\psi}_m\psi_n M^A\\
\widehat{\cal D}_m\Omega^A & = & \partial_m\Omega^A+\frac{1}{4}\widehat{\omega}_{mab}\gamma^{ab}\Omega^A-\frac{i}{2}\vec{V}_m\vec{\tau}\Omega^A+gf_{BC}{}^AB_m^B\Omega^C\\ & - & \frac{1}{4}\gamma^{ab}\psi_m\widehat{G}_{ab}^A + \frac{1}{2}\gamma^a\psi_m\widehat{\cal D}_a M^A - i\vec{\tau}\psi_m\vec{X}^A+\frac{1}{2\sqrt{3}}\gamma^{ab}\psi_m\widehat{F}_{ab} M^A.
\end{eqnarray*} 
The next step would be to construct an action for this multiplet. However, as will turn out and as expected from the four dimensional case \cite{Breitenlohner:1981ej}, this action is rather complicated. To circumvent its direct construction, it is useful to construct first the linear multiplet, for which an action is easily found. Since a certain combination of fields belonging to the Super Yang-Mills multiplet forms a linear multiplet, this formula can then be used to give an action for the Super Yang-Mills multiplet, as will be discussed at the end of the next section.

\section{The Linear Multiplet}\label{sec3}
 The linear multiplet is a useful device for the construction of invariant actions \cite{Breitenlohner:1980np}. Since the highest component of this  multiplet is a scalar $N$ which transforms in the globally supersymmetric case into a total derivative, an invariant is given by
 \[
 \sim\int d^5x~N.
 \]
 This relation is easily extended to the locally supersymmetric case. It will be shown below, following the construction in the four dimensional case \cite{Breitenlohner:1981ej}, that a certain combination of components of the minimal multiplet as constructed in \cite{Zucker:1999ej} forms a linear multiplet, so that knowing an action for the linear multiplet, an action for the minimal multiplet is known, too. Furthermore, the Super Yang-Mills multiplet can be embedded in a similar manner in the linear multiplet which will simplify the construction of an action for this multiplet considerably. 

The complete field content of the linear multiplet is given in table \ref{table2}.
\begin{table}
 \begin{center}
 \begin{tabular}{ccccccc}
 \hline\hline
 Type & Field & $SU(2)$ & \multicolumn{3}{c}{components}\\
&&&bosonic&&fermionic\\
 \hline
isotriplet & $\vec{Y}$ & \bf 3 & 3&& \\
fermion & $\rho$ & \bf 2 & && 8 & \\
scalar & $N$ & \bf 1 & 1 &&\\
vector & $W_a$ & \bf 1 & $5-1$&&\\
\hline\hline
&&&8&+&8\\
\end{tabular}
\end{center}
\caption{Field content of the linear multiplet in $D=5$. The subtracted degree of freedom of the vectorfield is due to the constraint (\ref{const}).}\label{table2}
 \end{table}
It contains a vector which is conserved in the rigid supersymmetric limit, i.e. $\partial\cdot W=0$. Of course, this constraint will be modified in the locally supersymmetric case; its actual form is given in eq. (\ref{const}). 

In the globally supersymmetric limit, the linear multiplet is dual to the Maxwell multiplet, i.e. by requiring invariance of the coupling term
\footnote{From this point of view the conservation of the vector $W_a$ is a direct consequence of the fact that $B_a$ is a gauge field.}
\[
\sim\int d^5x\left( B_aW^a+MN+i\bar{\Omega}\rho-\vec{X}\vec{Y}\right)
\]
we deduced the rigid transformation laws of the linear multiplet which were then completed to the locally supersymmetric ones by enforcing the gauge algebra (\ref{alg2}). 

In addition, the fields may transform in a real representation of a  nonabelian gauge group. 
If such a gauge symmetry is included, the fields tranform like
\[
\delta N=i\varepsilon^AT_A N,
\]
where the $T^A$ generate this representation and $\varepsilon^A$ is the transformation parameter. 
The supersymmetry transformation laws we find are
\begin{eqnarray*}
\delta \vec{Y} & = & \bar{\varepsilon} \vec{\tau}\rho\\
\delta \rho & = & \gamma^a\varepsilon W_a + \varepsilon N-\frac{i}{2} \gamma^a\vec{\tau}\varepsilon \widehat{\cal D}_a\vec{Y}+i\gamma^{ab}\vec{\tau}\varepsilon v_{ab}\vec{Y}\\ & + & 6i\vec{\tau}\varepsilon(\vec{t}\times\vec{Y})-\frac{i}{2\sqrt{3}}\gamma^{ab}\vec{\tau}\varepsilon\widehat{F}_{ab}\vec{Y}+\frac{g}{2}\vec{\tau}\varepsilon M\vec{Y}\\
\delta W^a & = & \frac{i}{2}\bar{\varepsilon}\gamma^{ab}\widehat{\cal D}_b\rho+\frac{1}{2}\bar{\varepsilon}\vec{\tau}\gamma^{abc}\widehat{\cal R}_{bc}\vec{Y}-\frac{i}{2}\bar{\varepsilon}\gamma^{abc}\rho v_{bc}+\frac{3i}{2}\bar{\varepsilon}\gamma_b\rho v^{ab}\\ & + & 4\bar{\varepsilon}\vec{\tau}\gamma^a\rho\vec{t}+\frac{i}{2\sqrt{3}}\bar{\varepsilon}\gamma^{abc}\rho \widehat{F}_{bc}-\frac{i}{2\sqrt{3}}\bar{\varepsilon}\gamma_b\rho \widehat{F}^{ab}+\frac{ig}{2}\bar{\varepsilon}\vec{\tau}\gamma^a\Omega\vec{Y}+\frac{g}{2}\bar{\varepsilon}\gamma^aM\rho\\
\delta N & = & -\frac{i}{2}\bar{\varepsilon}\gamma^a \widehat{\cal D}_a\rho-\frac{1}{2} \bar{\varepsilon}\vec{\tau}\gamma^{ab}\widehat{\cal R}_{ab}\vec{Y}-8\bar{\varepsilon}\vec{\tau}\lambda\vec{Y}-\frac{3i}{4}\bar{\varepsilon}\gamma^{ab}\rho v_{ab}-3\bar{\varepsilon}\vec{\tau}\rho\vec{t}-\frac{ig}{2}\bar{\varepsilon}\vec{\tau}\Omega\vec{Y}\\
\end{eqnarray*}
Here we use Lie-algebra valued expressions for the fields belonging to the Super Yang-Mills multiplet, e.g. $\Omega=\Omega^AT_A$.
Explicitly, the supercovariant derivatives appearing in the transformation laws are
\begin{eqnarray*}
\widehat{\cal D}_m \vec{Y} & = & \partial_m \vec{Y}+\vec{V}_m\times\vec{Y}-igB_m\vec{Y}-\bar{\rho}\vec{\tau}\psi_m\\
\widehat{\cal D}_m\rho & = & \partial_m \rho+\frac{1}{4}\widehat{\omega}_{mab}\gamma^{ab}\rho-\frac{i}{2}\vec{V}_m\vec{\tau}\rho-igB_m\rho - \gamma^a\psi_mW_a-\psi_mN+\frac{i}{2}\gamma^a\vec{\tau}\psi_m\widehat{\cal D}_a\vec{Y}\\ & - & i\gamma^{ab}\vec{\tau}\psi_m v_{ab}\vec{Y}-6i \vec{\tau}\psi_m(\vec{t}\times\vec{Y})+\frac{i}{2\sqrt{3}}\gamma^{ab}\vec{\tau}\psi_m\widehat{F}_{ab}\vec{Y}-\frac{g}{2}\vec{\tau}\psi_m M\vec{Y}\\
\widehat{\cal D}_mN & = & \partial_mN -igB_mN+ \frac{i}{2}\bar{\psi}_m\gamma^a \widehat{\cal D}_a\rho + \frac{1}{2} \bar{\psi}_m\vec{\tau}\gamma^{ab}\widehat{\cal R}_{ab}\vec{Y}\\ & + & \bar{\psi}_m\vec{\tau}\lambda\vec{Y}+\frac{3i}{4}\bar{\psi}_m\gamma^{ab}\rho v_{ab}+3\bar{\psi}_m\vec{\tau}\rho\vec{t}+\frac{ig}{2}\bar{\psi}_m\vec{\tau}\Omega\vec{Y}\\
\end{eqnarray*}
and the constraint which was mentioned above and which is required for closure of the algebra on $N$ and $W_a$ is
\begin{equation}
\widehat{\cal D}_aW^a+\frac{i}{2}\bar{\rho}\gamma^{ab}\widehat{\cal R}_{ab}-8ig M\vec{Y}\vec{t}+ig\vec{X}\vec{Y}+g\bar{\Omega}\rho-igMN=0\label{const}
\end{equation}
The supercovariant derivative on $W^a$ which appears in this expression is defined by
\begin{eqnarray*}
\widehat{\cal D}_m W^a & = & \partial_mW^a+\widehat{\omega}_m{}^{ab}W_b -igB_mW^a-\frac{i}{2}\bar{\psi}_m\gamma^{ab}\widehat{\cal D}_b\rho-\frac{1}{2}\bar{\psi}_m\gamma^{abc}\vec{\tau}\widehat{\cal R}_{bc}\vec{Y}\\ & + & \frac{i}{2}\bar{\psi}_m\gamma^{abc}\rho v_{bc} - \frac{3i}{2}\bar{\psi}_m\gamma_b\rho v^{ab} - 4\bar{\psi}_m\vec{\tau}\gamma^a\rho\vec{t}-\frac{i}{2\sqrt{3}}\bar{\psi}_m\gamma^{abc}\rho \widehat{F}_{bc}\\ & + & \frac{i}{2\sqrt{3}}\bar{\psi}_m\gamma_b\rho \widehat{F}^{ab}-\frac{ig}{2}\bar{\psi}_m\vec{\tau}\gamma^a\Omega\vec{Y}-\frac{g}{2}\bar{\psi}_m\gamma^aM\rho.
\end{eqnarray*}
We now give an invariant for the case of vanishing gauge group: $S=\int d^5x~e{\cal L}$, with a lagrangian
\begin{equation}
\begin{split}
{\cal L} =-4N+2i\bar{\psi}_a\gamma^a\rho- & 32\vec{t}\vec{Y}+\bar{\psi}_a\vec{\tau}\gamma^{ab}\psi_b\vec{Y}\\
 & -\frac{2}{\sqrt{3}}(4W^a+2i\bar{\rho}\gamma^{ab}\psi_b+\bar{\psi}_b\vec{\tau}\gamma^{abc}\psi_c\vec{Y})A_a\label{lag1}
\end{split}
\end{equation}
is invariant under supersymmetry and $U(1)$ gauge transformations. To check the latter symmetry the constraint (\ref{const}) has to be used. 
Note the formal similarity of this action with the four dimensional result \cite{Breitenlohner:1981ej}.

As stated above, a certain combination of fields belonging to the minimal multiplet forms a linear multiplet. The exact correspondence is
\[
(\vec{Y},\rho,N, W^a)=(\vec{t},\lambda, C, \frac{1}{4}\widehat{\cal D}_bv^{ab}+\frac{1}{48}\varepsilon^{abcde}\widehat{F}_{bc}\widehat{F}_{de}).
\]
Using this relation in (\ref{lag1}) leads to the action for the minimal multiplet given in \cite{Zucker:1999ej}.

As for the minimal multiplet it is also possible for the Super Yang-Mills multiplet of the preceding section to find certain combinations of fields which form the components of a linear multiplet. Since we know an action formula for this multiplet, eq. (\ref{lag1}), we can find in this way an action for the Super Yang-Mills multiplet. Because the exact correspondence is rather complicated, we give only the lowest dimensional field of the linear multiplet:
\begin{equation}
(\vec{Y}^{AB}, \rho^{AB}, N^{AB}, W_a^{AB}) = \left(\frac{1}{2}(M^A\vec{X}^B+M^B\vec{X}^A+\frac{1}{2}\bar{\Omega}^A\vec{\tau}\Omega^B-4 M^AM^B\vec{t}), \ldots\right).\label{emb1}
\end{equation}
The remaining components of this linear multiplet can then be computed by repeated supersymmetry variation of this expression.
Note again the similarity of this result with the result of Breitenlohner and Sohnius  \cite{Breitenlohner:1981ej}. 

The next step is to project out a gauge singlet by considering the contraction of the indices $A$ and $B$ in (\ref{emb1}). We write this as a trace. The lagrangian density for the Super Yang-Mills multiplet is then:
\begin{equation}
\begin{split}
{\cal L}_{YM} =  tr \bigg ( & - \frac{1}{4}\widehat{G}_{ab}\widehat{G}^{ab}+\frac{1}{2}\widehat{\cal D}_aM\widehat{\cal D}^aM+2\vec{X}\vec{X}+i\bar{\Omega}\gamma^m\widehat{\cal D}_m\Omega+ 64MM\vec{t}\vec{t}\\ & - 16 M\vec{X}\vec{t} + 2(\widehat{G}_{ab}-\frac{1}{\sqrt{3}}M\widehat{F}_{ab})(v^{ab} + \frac{1}{2\sqrt{3}}\widehat{F}^{ab})M + 8CM^2\\ & + \frac{1}{4\sqrt{3}}\varepsilon^{abcde}A_aG_{bc}G_{de} - 16i\bar{\lambda}\Omega M - \frac{i}{2}\bar{\Omega}\gamma^{ab}\Omega v_{ab} - 2\bar{\Omega}\vec{\tau}\Omega\vec{t}\\ & + g[\bar{\Omega},\Omega]M - \bar{\psi}_a\vec{\tau}\gamma^a\Omega\vec{X} + 8\bar{\Omega}\vec{\tau}\gamma^a\psi_a\vec{t}M -\frac{i}{2}\bar{\Omega}\gamma^a\gamma^b\psi_b\widehat{\cal D}_aM\\ & -\frac{i}{4}\bar{\psi}_a\gamma^a\gamma^{bc}\Omega\widehat{G}_{bc} + 4i\bar{\Omega}\gamma_a\psi_bv^{ab}M + \frac{i}{2\sqrt{3}}\bar{\psi}_a\gamma^a\gamma^{bc}\Omega\widehat{F}_{bc}M\\ & + ~ \frac{i}{4}\bar{\psi}_m\gamma^{mnab}\psi_n\widehat{G}_{ab}M - ~ 4i\bar{\psi}_a\gamma^a\lambda M^2- ~ 2\bar{\psi}_a\vec{\tau}\gamma^{ab}\psi_b\vec{t}M^2\\ & - \frac{i}{4\sqrt{3}}\bar{\psi}_m\gamma^{mnab}\psi_n\widehat{F}_{ab}M^2 - iv^{ab}\bar{\psi}_a\psi_bM^2   + \frac{1}{4}\bar{\psi}_a\vec{\tau}\gamma^{ab}\psi_b\bar{\Omega}\vec{\tau}\Omega\\ & + \frac{1}{4}\bar{\Omega}\gamma^{ab}\Omega\bar{\psi}_a\psi_b - \frac{1}{2}\bar{\psi}_n\gamma^a\psi_m\bar{\Omega}\gamma^{mn}\psi_aM + \bar{\psi}_p\gamma^p\psi_m\bar{\Omega}\gamma^{mn}\psi_nM\\ & - \bar{\Omega}\gamma^{mn}\psi_n\bar{\Omega}\psi_m + \frac{1}{2}\bar{\Omega}\gamma^{mnp}\psi_n\bar{\psi}_m\psi_pM + \frac{1}{8}\bar{\psi}_m\gamma^{mnpq}\psi_n\bar{\psi}_p\psi_qM^2\bigg)
\end{split}\label{symlag}
\end{equation} 
Note the appearance of a term linear in $C$ (the last term in the second line), as expected from the counting of degrees of freedom. As in the pure supergravity case, this field has to be replaced using the nonlinear multiplet.
\section{Gauged Supergravity\label{sec4}}
As discussed in the introduction, there are two possibilities for gauged supergravity. First, one can gauge the complete automorphisms by introducing in addition to the supergravity multiplet an $SU(2)$ vector multiplet. The second posssibility is to gauge an $SO(2)$ subgroup of the $SU(2)$, where the gauge field of the $SO(2)$ is the graviphoton, which belongs to the supergravity multiplet. We will discuss the first possibility only roughly, following the pioneering work of de Wit, \cite{deWit:1980ib}. For the second possibility , i.e. {\it AdS} supergravity, we will present detailed formulas. Our strategy in this case  follows closely the work of de Wit {\it et al.} on four dimensional conformal supergravity, \cite{deWit:1981tn}.

To gauge the automorphism group of the supersymmetry algebra we have to return to the nonlinear multiplet which has been constructed in \cite{Zucker:1999ej}. Its components are listed in table \ref{nonl}.
\begin{table}
\begin{center}
\begin{tabular}{ccccccc}
\hline\hline
Type & Field & $SU(2)\times SU(2)'$ & dimension & \multicolumn{3}{c}{components}\\
&&&&bosonic&&fermionic\\
\hline
scalar & $\phi^i{}_\alpha$ & $({\mathbf 2},{\mathbf 2})$ & 0 & 3\\
spinor &  $\chi$ & $({\mathbf 2},{\mathbf 1})$ & 2 &&& 8 \\
scalar & $\varphi$ &$({\mathbf 1},{\mathbf 1})$ & 5/2 & 1 \\
vector & $V_a$ & $({\mathbf 1},{\mathbf 1})$ & 5/2 & $5-1$ \\
\hline\hline
&&&&8&+&8\\
\end{tabular}
\end{center}
\caption{Field content of the nonlinear multiplet. The subtracted degree of freedom of $V_a$ is due to the constraint (\ref{const2}).}\label{nonl}
\end{table}
A constraint of the form $\partial_aV^a+\ldots=0$ was required to render this multiplet supersymmetric.
The index $\alpha$ of $\phi^i{}_\alpha$ transformed under global $SU(2)'$ transformations. The first  step towards gauged supergravity is to gauge this $SU(2)'$ by introducing a Super Yang-Mills multiplet with components 
\begin{equation}
(\vec{B}_m, \vec{M}, \vec{\Omega}, \vec{\widetilde{X}}),\label{gaugemultiplet}
\end{equation}
where the arrows denote the $\mathbf{3}$ of $SU(2)'$ and the tilde on $\vec{\widetilde{X}}$ denotes the $\mathbf{3}$ of the automorphism group $SU(2)$. To gauge the $SU(2)'$, the derivatives acting on $\phi^i{}_\alpha$ which appear in various transformation laws of fields belonging to the nonlinear multiplet have to be replaced by gauge covariant derivatives:
\begin{equation}
\partial_m\phi^i{}_\alpha \to \partial_m\phi^i{}_\alpha+\frac{ig}{2}\phi^i{}_\beta(\vec{\tau})^\beta{}_\alpha\vec{B}_m.\label{der1} 
\end{equation}
However, imposing the commutator algebra with the modification due to the extra $SU(2)'$ given in eq. (\ref{alg2}) on the nonlinear multiplet requires further modifications of the transformation laws. These are (besides the implicit replacements of the derivatives by covariant derivatives, cf. (\ref{der1}), and various supercovariantizations):
\begin{eqnarray*}
\Delta \chi^i & = & \frac{-ig}{4}\phi^i{}_\alpha(\vec{\tau})^\alpha{}_\beta\phi^\beta{}_j\varepsilon^j\vec{M}\\
\Delta\varphi & = & \frac{g}{2}\bar{\varepsilon}_i\chi^j\phi^i{}_\alpha(\vec{\tau})^\alpha{}_\beta\phi^\beta{}_j\vec{M}+\frac{g}{2}\bar{\varepsilon}_i\vec{\Omega}^j\phi^i{}_\alpha(\vec{\tau})^\alpha{}_\beta\phi^\beta{}_j\\
\Delta V_a & = & \frac{g}{2}\bar{\varepsilon}_i\gamma_a\chi^j\phi^i{}_\alpha(\vec{\tau})^\alpha{}_\beta\phi^\beta{}_j\vec{M}+\frac{g}{2}\bar{\varepsilon}_i\gamma_a\vec{\Omega}^j\phi^i{}_\alpha(\vec{\tau})^\alpha{}_\beta\phi^\beta{}_j,
\end{eqnarray*}
so that $\delta^{new}=\delta^{old}+\Delta$.
Furthermore, the constraint of the nonlinear multiplet as given in \cite{Zucker:1999ej} is modified. Besides the implicit modifications due to covariantizations and supercovariantizations there appear also extra terms:
\begin{equation}
\begin{split}
C & + \ldots -\frac{g}{8}\vec{\widetilde{X}}(\tilde{\tau})^j{}_i\phi^i{}_\alpha(\vec{\tau})^\alpha{}_\beta\phi^\beta{}_j\\ & -  \frac{g}{4}\bar{\chi}_i(\chi^j\vec{M}+2\vec{\Omega}^j)\phi^i{}_\alpha(\vec{\tau})^\alpha{}_\beta\phi^\beta{}_j
+\frac{g}{4}(\tilde{\tau})^j{}_i\phi^i{}_\alpha(\vec{\tau})^\alpha{}_\beta\phi^\beta{}_j\vec{M}\tilde{t}+\frac{g^2}{16}\vec{M}^2=0.
\end{split}\label{const2}
\end{equation}
The dots represent again the terms already present in \cite{Zucker:1999ej} and for clarity we have given the $C$--term.

To gauge the complete $SU(2)$ automorphism group with a propagating field one should proceed as follows. First, one has to plug in the modified constraint (\ref{const2}) in the Super Yang-Mills lagrangian (\ref{symlag}). Next, the $SU(2)\times SU(2)'$ is broken to a residual $SU(2)$ by choosing
\begin{equation}
\phi^i{}_\alpha=\delta^i_\alpha.\label{fix}
\end{equation}
To maintain this gauge, a field dependent $SU(2)$ transformation with parameter
\[
\vec{\xi}=2\bar{\varepsilon}\vec{\tau}\chi+\vec{\eta}
\]
has to be added to the transformation laws of the fields. Here $\vec{\eta}$ is the transformation parameter of the $SU(2)'$. As a consequence, there are now two fields, $\vec{B}_m$ and $\vec{V}_m$, transforming as gauge fields of the residual $SU(2)$. Further, there is a derivative term $\sim \partial\varepsilon$ in the supersymmetry transformation law of $\vec{V}_m$. All this can be remedied by redefining
\begin{equation}
\vec{B}_m'=  \vec{V}_m-g\vec{B}_m+2\bar{\psi}_m\vec{\tau}\chi
\end{equation}
$\vec{B}_m'$ is now inert under gauge transformations and of auxilliary dimension.

To gauge an $SO(2)$ subgroup of the $SU(2)$ automorphism group one could proceed by setting
\begin{equation}
\vec{\Omega}=\vec{\widetilde{X}}=0,\qquad \vec{M}=\frac{1}{\sqrt{2}}(0,1,0)^T,\qquad \vec{B}_m=\sqrt{\frac{2}{3}}(0,A_m,0)^T.\label{fix1}
\end{equation}
Inspection of the supersymmetry transformation laws for the Super Yang-Mills multiplet (\ref{sym}) shows that this choice is supersymmetric. Furthermore, one should perform the following redefinitions which lead to canonical kinetic terms:
\begin{equation}
\lambda'=\lambda+\frac{1}{4}\gamma^{ab}\widehat{\cal R}_{ab},\qquad v'_{ab}=v_{ab}-\frac{1}{2\sqrt{3}}\widehat{F}_{ab}.\label{redef1}
\end{equation}
However, a shortcut of this procedure is to start by imposing (\ref{fix1}) and the redefinitions (\ref{redef1}) first. After having replaced the $C$ field in the lagrangian by the constraint (\ref{const2}) and having chosen the gauge (\ref{fix}), the original $SU(2)\times SU(2)'$ is broken to $SO(2)$. To stay in the gauge (\ref{fix}), an $SU(2)$ transformation with parameter
\[
\vec{\xi}=2\bar{\varepsilon}\vec{\tau}\chi+(0,\eta^2,0)^T
\]
has to be addded to the supersymmetry transformation laws. Again, as in the case with gauged $SU(2)$ discussed above, there are now two fields which transform as gauge fields for the $SO(2)$, namely $V^2_m$ and $A_m$. To get rid of this, we define a new vectorfield
\[
\vec{V}_m'= \vec{V}_m-g\sqrt{\frac{2}{3}}(0,A_m,0)^T-2\bar{\psi}_m\vec{\tau}\chi
\] 
which is inert under the $SO(2)$. 
After replacing
\[
\vec{t}\to \vec{t}+\frac{g}{12\sqrt{2}}(0,1,0)^T
\]
the lagrangian becomes
\[
{\cal L}={\cal L}_1+{\cal L}_2+{\cal L}_3+\frac{g^{'2}}{\kappa^4}-\sqrt{3}\frac{g'}{\kappa}\bar{\chi}\tau^2\chi-\frac{\sqrt{3}g'}{4\kappa}\bar{\psi}_a\tau^2\gamma^{ab}\psi_b
\]
where we have restored the gravitational coupling constant $\kappa$.
Here ${\cal L}_1+{\cal L}_2+{\cal L}_3$ are the lagrangians of the ungauged theory given in \cite{Zucker:1999ej}, with the derivatives acting on the spinor doublets $\psi_m$ and $\chi$ replaced by gauge covariant derivatives,
\[
\partial_m\to \partial_m-ig'A_m\tau^2
\]
and we have defined a new gauge coupling constant $g'=g/\sqrt{6}$. On-shell only
\[
{\cal L}={\cal L}_1+\frac{g^{'2}}{\kappa^4}-\frac{\sqrt{3}g'}{4\kappa}\bar{\psi}_a\tau^2\gamma^{ab}\psi_b
\]
remains. This on-shell lagrangian agrees up to conventions with a special case of the more general theory discussed in \cite{Gunaydin:1985ak}. 

\medskip
This work was partially supported by the European Commission programs ERBFMRX-CT96-0045 and CT96-0090.

\end{document}